\documentclass[12pt]{article}
\usepackage{axodraw}

\catcode`@=12
\begin{document}
\thispagestyle{empty}
\begin{flushright} 
UCRHEP-T344\\ 
August 2002\
\end{flushright}
\vspace{0.5in}
\begin{center}
{\LARGE	\bf Neutrino Mass Matrix a la Mode\\}
\vspace{1.2in}
{\bf Ernest Ma\\}
\vspace{0.2in}
{\sl Physics Department, University of California, Riverside, 
California 92521\\}
\vspace{1.2in}
\end{center}
\begin{abstract}\
With the accumulation of many years of solar and atmospheric neutrino 
oscillation data, the approximate form of the $3 \times 3$ neutrino mixing 
matrix is now known.  What is not known is the (presumably Majorana) neutrino 
mass matrix ${\cal M}_\nu$ itself.  In this chapter, the approximate form of 
${\cal M}_\nu$ is derived, leading to seven possible neutrino mass patterns: 
three have the normal hierarchy, two have the inverse hierarchy, and two 
have three nearly degenerate masses.  The generalization of this to allow 
$U_{e3} \neq 0$ with maximal CP violation is also discussed.  A specific 
automatic realization of this ${\cal M}_\nu$ from radiative corrections of 
an underlying non-Abelian discrete $A_4$ symmetry in the context of softly 
broken supersymmetry is presented.
\end{abstract}
\vspace{0.5in}
\noindent ------------------

\noindent Contribution to a special issue of the Proceedings of the Indian 
National Science Academy
\newpage
\baselineskip 24pt

With the recent addition of the SNO (Sudbury Neutrino Observatory) 
neutral-current data \cite{sno}, the overall picture of solar neutrino 
oscillations \cite{sol} is becoming quite clear.  Together with the 
well-established atmospheric neutrino data \cite{atm}, the $3 \times 3$ 
neutrino mixing matrix is now determined to a very good first approximation by
\begin{equation}
\pmatrix {\nu_e \cr \nu_\mu \cr \nu_\tau} = \pmatrix {\cos \theta & -\sin 
\theta & 0 \cr \sin \theta/\sqrt 2 & \cos \theta/\sqrt 2 & -1/\sqrt 2 \cr 
\sin \theta/\sqrt 2 & \cos \theta/\sqrt 2 & 1/\sqrt 2} \pmatrix {\nu_1 \cr 
\nu_2 \cr \nu_3},
\end{equation}
where $\nu_{1,2,3}$ are assumed to be Majorana neutrino mass eigenstates.  
In the above, $\sin^2 2 \theta_{atm} = 1$ is already assumed and $\theta$ 
is the solar mixing angle which is now known to be large but not maximal 
\cite{many}, i.e. $\tan^2 \theta \simeq 0.4$.  The $U_{e3}$ entry has been 
assumed zero but it is only required experimentally to be small \cite{react}, 
i.e. $|U_{e3}| < 0.16$. 

Denoting the masses of $\nu_{1,2,3}$ as $m_{1,2,3}$, the solar neutrino data 
\cite{sno,sol} require that $m_2^2 > m_1^2$ with $\theta < \pi/4$, and in 
the case of the favored large-mixing-angle solution \cite{many},
\begin{equation}
\Delta m^2_{sol} = m_2^2 - m_1^2 \simeq 5 \times 10^{-5}~{\rm eV}^2.
\end{equation}
The atmospheric neutrino data \cite{atm} require
\begin{equation}
|m_3^2 - m_{1,2}^2| \simeq 2.5 \times 10^{-3}~{\rm eV}^2,
\end{equation}
without deciding whether $m_3^2 > m_{1,2}^2$ or $m_3^2 < m_{1,2}^2$.

The big question now is what the neutrino mass matrix itself should look like. 
Of course it may be obtained by using Eq.~(1), i.e.
\begin{equation}
{\cal M}_\nu = \pmatrix {c^2 m_1 + s^2 m_2 & sc(m_1-m_2)/\sqrt 2 & sc(m_1-m_2)
/\sqrt 2 \cr sc(m_1-m_2)/\sqrt 2 & (s^2 m_1 + c^2 m_2 + m_3)/2 & (s^2 m_1 + 
c^2 m_2 - m_3)/2 \cr sc(m_1-m_2)/\sqrt 2 & (s^2 m_1 + c^2 m_2 - m_3)/2 & 
(s^2 m_1 + c^2 m_2 + m_3)/2},
\end{equation}
where $c \equiv \cos \theta$ and $s \equiv \sin \theta$.  However this is not 
very illuminating theoretically.  Instead it has been proposed \cite{ma02} 
that it be rewritten in the form
\begin{equation}
{\cal M}_\nu = \pmatrix {a+2b+2c & d & d \cr d & b & a+b \cr d & a+b & b}.
\end{equation}
To satisfy $m_2^2 > m_1^2$ for $\theta < \pi/4$, there are 2 cases to be 
considered.
 
\noindent (I) For $a+2b+c > 0$ and $c < 0$,
\begin{eqnarray}
m_1 &=& a+2b+c - \sqrt {c^2 + 2d^2}, \\ 
m_2 &=& a+2b+c + \sqrt {c^2 + 2d^2}, \\ 
m_3 &=& -a,\\
\tan \theta &=& \sqrt {2} d/ (\sqrt {c^2 + 2d^2} - c).
\end{eqnarray}
(II) For $a+2b+c < 0$ and $c > 0$,
\begin{eqnarray}
m_1 &=& a+2b+c + \sqrt {c^2 + 2d^2}, \\ 
m_2 &=& a+2b+c - \sqrt {c^2 + 2d^2}, \\ 
m_3 &=& -a,\\
\tan \theta &=& (\sqrt {c^2 + 2d^2} - c)/ \sqrt {2} d.
\end{eqnarray}

Note that $\theta$ depends only on the ratio $d/c$, which must be of order 
unity.  This shows the advantage for adopting the parametrization of 
Eq.~(5).  The constraints of Eqs.~(2) and (3) are then realized by the 
following 7 different conditions on $a$, $b$, and $c$.

(1) $||a+2b+c|-\sqrt{c^2+2d^2}| << |a+2b+c| << |a|$, ~~i.e. $|m_1| << |m_2| << 
|m_3|$.

(2) $\sqrt{c^2+2d^2} << |a+2b+c| << |a|$, ~~i.e. $|m_1| \simeq |m_2| << |m_3|$.

(3) $|a+2b+c| << \sqrt{c^2+2d^2} << |a|$, ~~i.e. $|m_1| \simeq |m_2| << |m_3|$.

(4) $|a|,~\sqrt{c^2+2d^2} << |a+2b+c|$, ~~i.e. $|m_3| << |m_1| \simeq |m_2|$.

(5) $|a|,~|a+2b+c| << \sqrt{c^2+2d^2}$, ~~i.e. $|m_3| << |m_1| \simeq |m_2|$.

(6) $\sqrt{c^2+2d^2} << ||a+2b+c|-|a|| << |a|$, ~~i.e. $|m_1| \simeq |m_2| 
\simeq |m_3|$.

(7) $|a+2b+c| << \sqrt{c^2+2d^2} \simeq |a|$, ~~i.e. $|m_1| \simeq |m_2| 
\simeq |m_3|$.

\noindent Cases (1) to (3) have the normal hierarchy. Cases (4) and (5) have 
the inverse hierarchy. Cases (6) and (7) have 3 nearly degenerate masses. 
The versatility of Eq.~(5) has clearly been demonstrated.

The above 7 cases encompass all models of the neutrino mass matrix that 
have ever been proposed which also satisfy Eq.~(1).  They are also very 
useful for discussing the possibility of neutrinoless double beta ($\beta 
\beta_{0\nu}$) decay in the context of neutrino oscillations \cite{bbmass}.  
The effective mass $m_0$ measured in $\beta \beta_{0\nu}$ decay is 
$|a+2b+2c|$.  However, neutrino oscillations constrain $|a+2b+c|$ and 
$\sqrt{c^2+2d^2}$, as well as $|d/c|$.  Using
\begin{equation}
|a+2b+2c| = ||a+2b+c| \pm |c|| = ||a+2b+c| \pm \cos 2 \theta \sqrt{c^2+2d^2}|,
\end{equation}
the following conditions on $m_0$ are easily obtained:
\begin{eqnarray}
(1) && m_0 \simeq \sin^2 \theta |m_2| \simeq \sin^2 \theta \sqrt 
{\Delta m^2_{sol}}, \\ 
(2) && m_0 \simeq |m_{1,2}| << \sqrt {\Delta m^2_{atm}}, \\ 
(3) && m_0 \simeq  \cos 2 \theta |m_{1,2}| << \cos 2 \theta \sqrt 
{\Delta m^2_{atm}}, \\ 
(4) && m_0 \simeq \sqrt {\Delta m^2_{atm}}, \\ 
(5) && m_0 \simeq \cos 2 \theta \sqrt {\Delta m^2_{atm}}, \\ 
(6) && m_0 \simeq |m_{1,2,3}|, \\ 
(7) && m_0 \simeq \cos 2 \theta |m_{1,2,3}|.
\end{eqnarray}
If $m_0$ is measured \cite{klapdor} to be significantly larger than 0.05 eV, 
then only Cases (6) and (7) are allowed.  However, as Eqs.~(20) and (21) 
show, the true mass of the three neutrinos is still subject to a two-fold 
ambiguity, which is a well-known result.

The underlying symmetry of Eq.~(5) which results in $U_{e3} = 0$ is its 
invariance under the interchange of $\nu_\mu$ and $\nu_\tau$.  Its mass 
eigenstates are then separated according to whether they are even 
$(\nu_{1,2})$ or odd $(\nu_3)$ under this interchange, as shown by Eq.~(1). 
To obtain $U_{e3} \neq 0$, this symmetry has to be broken.  One interesting 
possibility is to rewrite Eq.~(5) as
\begin{equation}
{\cal M}_\nu = \pmatrix {a+2b+2c & d & d^* \cr d & b & a+b \cr d^* & a+b & b},
\end{equation}
where $a,b,c$ are real but $d$ is complex.  This reduces to Eq.~(4) if 
$Im d = 0$, but if $Im d \neq 0$, then $U_{e3} \neq 0$.

To obtain $U_{e3}$ in a general way, first rotate to the basis spanned by 
$\nu_e, (\nu_\mu+\nu_\tau)/\sqrt 2$, and $(\nu_\tau-\nu_\mu)/\sqrt 2$, i.e.
\begin{equation}
{\cal M}_\nu = \pmatrix {a+2b+2c & \sqrt2 Red & -\sqrt2 i Imd \cr \sqrt2 Red 
& a+2b & 0 \cr -\sqrt 2 i Imd & 0 & -a}
\end{equation} 
Whereas ${\cal M}_\nu$ is diagonalized by
\begin{equation}
U {\cal M}_\nu U^T = \pmatrix {m_1 & 0 & 0 \cr 0 & m_2 & 0 \cr 0 & 0 & m_3},
\end{equation}
${\cal M}_\nu {\cal M}_\nu^\dagger$ is diagonalized by
\begin{equation}
U ({\cal M}_\nu {\cal M}_\nu^\dagger) U^\dagger = \pmatrix {|m_1|^2 & 0 & 0 
\cr 0 & |m_2|^2 & 0 \cr 0 & 0 & |m_3|^2}.
\end{equation}
Here
\begin{equation}
{\cal M}_\nu {\cal M}_\nu^\dagger = \pmatrix {(a+2b+2c)^2 + 2|d|^2 & 
2 \sqrt2 (a+2b+c) Red & 2 \sqrt2 i (a+b+c) Imd \cr 2 \sqrt2 (a+2b+c) Red & 
(a+2b)^2 + 2(Red)^2 & 2 i Red Imd \cr -2 \sqrt2 i (a+b+c) Imd & -2 i Red Imd 
& a^2 + 2(Imd)^2}.
\end{equation}
To obtain $U_{e3}$ for small $Imd$, consider the matrix
\begin{equation}
A = {\cal M}_\nu {\cal M}_\nu^\dagger - [a^2 + 2(Imd)^2] I,
\end{equation}
where $I$ is the identity matrix. Now $A$ is diagonalized by $U$ as well and 
$U_{e3}$ is simply given by
\begin{equation}
U_{e3} \simeq {A_{e3} \over A_{ee}} = {2 \sqrt 2 i (a+b+c) Imd \over 
(a+2b+2c)^2 - a^2 + 2 (Red)^2}
\end{equation}
to a very good approximation and leads to
\begin{eqnarray}
(1), (2), (3) && U_{e3} \simeq {-\sqrt 2 i Imd \over a}, \\ 
(4), (6) && U_{e3} \simeq {i Imd \over \sqrt 2 b}, \\ 
(5) && U_{e3} \simeq {\sqrt2 i Imd \over c}, \\ 
(7) && U_{e3} \simeq {\sqrt2 i (a+c) Imd \over c^2 - a^2 + 2(Red)^2}.
\end{eqnarray}
In all cases, the magnitude of $U_{e3}$ can be as large as the present 
experimental limit \cite{react} of 0.16 and its phase is $\pm \pi/2$. 
Thus the CP violating effect in neutrino oscillations is predicted 
to be maximal by Eq.~(22), which is a very desirable scenario for future 
long-baseline neutrino experiments.

The above analysis shows that for $U_{e3} = 0$ and $\sin^2 2 \theta_{atm} = 1$,
the seven cases considered cover all possible patterns of the $3 \times 3$ 
Majorana neutrino mass matrix, as indicated by present atmospheric and solar 
neutrino data.  Any successful model should predict Eq.~(5) at least as a 
first approximation.  One such example already exists \cite{a4}, where 
$b=c=d=0$ corresponds to the non-Abelian discrete symmetry $A_4$, i.e. the 
finite group of the rotations of a regular tetrahedron.  This leads to 
Case (6), i.e. three nearly degenerate masses, with the common mass equal 
to that measured in $\beta \beta_{0\nu}$ decay.  It has also been shown 
recently \cite{bmv} that starting with this pattern, the correct mass 
matrix, i.e. Eq.~(22) with the complex phase in the right place, is 
automatically obtained with the most general application of radiative 
corrections.  In particular, if soft supersymmetry breaking is assumed to 
be the origin of these radiative corrections, then the neutrino mass matrix 
is correlated with flavor violation in the slepton sector, and may be tested 
in future collider experiments.

Suppose that at some high energy scale, the charged lepton mass matrix and 
the Majorana neutrino mass matrix are such that after diagonalizing the 
former, i.e.
\begin{equation}
{\cal M}_l = \pmatrix {m_e & 0 & 0 \cr 0 & m_\mu & 0 \cr 0 & 0 & m_\tau},
\end{equation}
the latter is of the form
\begin{equation}
{\cal M}_\nu = \pmatrix {m_0 & 0 & 0 \cr 0 & 0 & m_0 \cr 0 & m_0 & 0}.
\end{equation}
From the high scale to the electroweak scale, one-loop radiative corrections 
will change ${\cal M}_\nu$ as follows:
\begin{equation}
({\cal M}_\nu)_{ij} \to ({\cal M}_\nu)_{ij} + R_{ik} ({\cal M}_\nu)_{kj} 
+ ({\cal M}_\nu)_{ik} R_{kj},
\end{equation}
where the radiative correction matrix is assumed to be of the most general 
form, i.e.
\begin{equation}
R = \pmatrix {r_{ee} & r_{e\mu} & r_{e\tau} \cr r_{e\mu}^* & r_{\mu\mu} & 
r_{\mu\tau} \cr r_{e\tau}^* & r_{\mu\tau}^* & r_{\tau\tau}}.
\end{equation}
Thus the observed neutrino mass matrix is given by
\begin{equation}
{\cal M}_\nu = m_0 \pmatrix {1+2r_{ee} & r_{e\tau} + r_{e\mu}^* & r_{e\mu} + 
r_{e\tau}^* \cr r_{e\mu}^* + r_{e\tau} & 2r_{\mu\tau} & 1+r_{\mu\mu}+
r_{\tau\tau} \cr r_{e\tau}^* + r_{e\mu} & 1+r_{\mu\mu}+r_{\tau\tau} & 
2r_{\mu\tau}^*}.
\end{equation}
Now $r_{\mu\tau}$ may be chosen real by absorbing its phase into $\nu_\mu$ 
and $\nu_\tau$.  Then using the redefinitions:
\begin{eqnarray}
&& \delta_0 \equiv r_{\mu\mu} + r_{\tau\tau} - 2r_{\mu\tau}, \\ 
&& \delta \equiv 2r_{\mu\tau}, \\
&& \delta' \equiv r_{ee} - {1 \over 2} r_{\mu\mu} - {1 \over 2} r_{\tau\tau} 
- r_{\mu\tau}, \\
&& \delta'' \equiv r_{e\mu}^* + r_{e\tau},
\end{eqnarray}
the neutrino mass matrix becomes
\begin{equation}
{\cal M}_\nu = m_0 \pmatrix{1+\delta_0+2\delta+2\delta' & \delta'' & 
\delta''^* \cr \delta'' & \delta & 1+\delta_0+\delta \cr \delta''^* & 
1+\delta_0+\delta & \delta},
\end{equation}
which is exactly that of Eq.~(22). In other words, starting with Eq.~(34), 
the correct ${\cal M}_\nu$ is automatically obtained.  [To simplify Eq.~(42) 
without any loss of generality, $\delta_0$ will be set equal to zero from 
here on.]

The successful derivation of Eq.~(42) depends on having Eqs.~(33) and (34). 
To be sensible theoretically, they should be maintained by a symmetry. 
At first sight, it appears impossible that there can be a symmetry which 
allows them to coexist.  Here is where the non-Abelian discrete symmetry 
$A_4$ comes into play \cite{a4}.  The key is that $A_4$ has three inequivalent 
one-dimensional representations \underline {1}, \underline {1}$'$, 
\underline {1}$''$, and one three-dimensional reprsentation \underline {3}, 
with the decomposition
\begin{equation}
\underline {3} \times \underline {3} = \underline {1} + \underline {1}' + 
\underline {1}'' + \underline {3} + \underline {3}.
\end{equation}
This allows the following natural assignments of quarks and leptons:
\begin{eqnarray}
&& (u_i,d_i)_L, ~~ (\nu_i,e_i)_L \sim \underline {3}, \\ 
&& u_{1R}, ~~ d_{1R}, ~~ e_{1R} \sim \underline {1}, \\ 
&& u_{2R}, ~~ d_{2R}, ~~ e_{2R} \sim \underline {1}', \\ 
&& u_{3R}, ~~ d_{3R}, ~~ e_{3R} \sim \underline {1}''.
\end{eqnarray}
Heavy fermion singlets are then added \cite{bmv}:
\begin{equation}
U_{iL(R)}, ~~ D_{iL(R)}, ~~ E_{iL(R)}, ~~ N_{iR} \sim \underline {3},
\end{equation}
together with the usual Higgs doublet and new heavy singlets:
\begin{equation}
(\phi^+,\phi^0) \sim \underline {1}, ~~~~ \chi^0_i \sim \underline {3}.
\end{equation}
With this structure, charged leptons acquire an effective Yukawa coupling 
matrix $\bar e_{iL} e_{jR} \phi^0$ which has 3 arbitrary eigenvalues 
(because of the 3 independent couplings to the 3 inequivalent one-dimensional 
representations) and for the case of equal vacuum expectation values of 
$\chi_i$, i.e.
\begin{equation}
\langle \chi_1 \rangle = \langle \chi_2 \rangle = \langle \chi_3 \rangle = u,
\end{equation}
the unitary transformation $U_L$ which diagonalizes ${\cal M}_l$ is given by
\begin{equation}
U_L = {1 \over \sqrt 3} \pmatrix {1 & 1 & 1 \cr 1 & \omega & \omega^2 \cr 
1 & \omega^2 & \omega},
\end{equation}
where $\omega = e^{2\pi i/3}$.  This implies that the effective neutrino 
mass operator, i.e. $\nu_i \nu_j \phi^0 \phi^0$, is proportional to
\begin{equation}
U_L^T U_L = \pmatrix {1 & 0 & 0 \cr 0 & 0 & 1 \cr 0 & 1 & 0},
\end{equation}
exactly as desired \cite{a4,bmv}.

To derive Eq.~(52), the validity of Eq.~(50) has to be proved.  This is 
naturally accomplished in the context of supersymmetry.  Let $\hat \chi_i$ 
be superfields, then its superpotential is given by
\begin{equation}
\hat W = {1 \over 2} M_\chi (\hat \chi_1 \hat \chi_1 + \hat \chi_2 \hat \chi_2 
+ \hat \chi_3 \hat \chi_3) + h_\chi \hat \chi_1 \hat \chi_2 \hat \chi_3.
\end{equation}
Note that the $h_\chi$ term is invariant under $A_4$, a property not found 
in $SU(2)$ or $SU(3)$.  The resulting scalar potential is
\begin{equation}
V = |M_\chi \chi_1 + h_\chi \chi_2 \chi_3|^2 + |M_\chi \chi_2 + h_\chi \chi_3 
\chi_1|^2 + |M_\chi \chi_3 + h_\chi \chi_1 \chi_2|^2.
\end{equation}
Thus a supersymmetric vacuum $(V=0)$ exists for
\begin{equation}
\langle \chi_1 \rangle = \langle \chi_2 \rangle = \langle \chi_3 \rangle = u 
= -M_\chi /h_\chi,
\end{equation}
proving Eq.~(50), with the important additional result that the spontaneous 
breaking of $A_4$ at the high scale $u$ does not break the supersymmetry.

\begin{figure}
\begin{center}
\begin{picture}(270,110)(0,0)
\ArrowLine(0,50)(50,50)
\ArrowLine(270,50)(220,50)
\Line(50,50)(170,50)
\ArrowLine(170,50)(220,50)
\DashCArc(110,50)(60,0,180){4}
\Text(110,40)[]{$\tilde w$}
\Text(25,40)[]{$\nu_\mu$}
\Text(195,60)[]{$\nu_\tau$}
\Text(245,60)[]{$\nu_\mu$}
\Text(55,105)[]{$\tilde \mu_L$}
\Text(165,105)[]{$\tilde \tau_L$}
\Text(110,110)[]{$\times$}
\DashArrowLine(170,10)(220,50){4}
\DashArrowLine(270,10)(220,50){4}
\Text(170,0)[]{$\phi_2^0$}
\Text(270,0)[]{$\phi_2^0$}
\end{picture}
\end{center}
\caption{Wavefunction contribution to $\delta$ in supersymmetry.}
\end{figure}
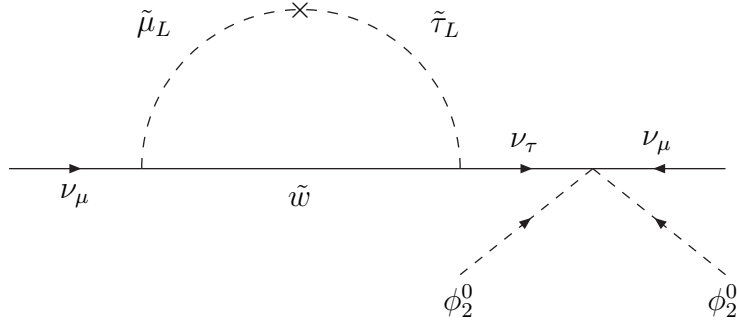

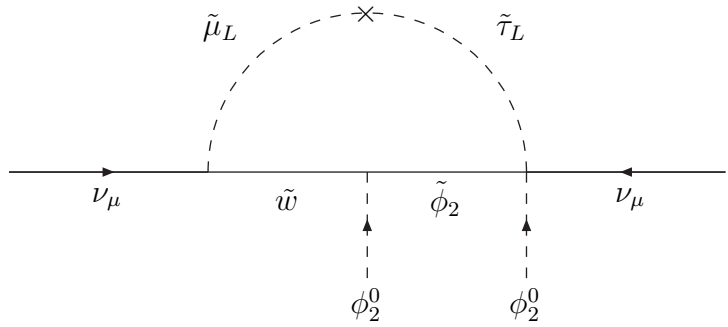
\begin{figure}
\begin{center}
\begin{picture}(270,110)(0,0)
\ArrowLine(0,50)(75,50)
\ArrowLine(270,50)(195,50)
\Line(0,50)(270,50)
\DashArrowLine(135,10)(135,50){4}
\DashArrowLine(195,10)(195,50){4}
\DashCArc(135,50)(60,0,180){4}
\Text(105,40)[]{$\tilde w$}
\Text(165,40)[]{$\tilde \phi_2$}
\Text(37,40)[]{$\nu_\mu$}
\Text(235,40)[]{$\nu_\mu$}
\Text(135,0)[]{$\phi_2^0$}
\Text(195,0)[]{$\phi_2^0$}
\Text(80,105)[]{$\tilde \mu_L$}
\Text(190,105)[]{$\tilde \tau_L$}
\Text(135,110)[]{$\times$}
\end{picture}
\end{center}
\caption{Vertex contribution to $\delta$ in supersymmetry.}
\end{figure}

To generate the proper radiative corrections which will result in a 
realistic Majorana neutrino mass matrix, $A_4$ is assumed broken also 
by the soft supersymmetry breaking terms.  In particular, the mass-squared 
matrix of the left sleptons will be assumed to be arbitrary.  This allows 
$r_{\mu\tau}$ to be nonzero through $\tilde \mu_L - \tilde \tau_L$ mixing, 
from which the parameter $\delta$ may be evaluated, as shown in Figs. 1 and 
2.  For illustration, using the approximation that $\tilde m_1^2 >> \mu^2 >> 
M_{1,2}^2 = \tilde m_2^2$, where $\mu$ is the Higgsino mass and $M_{1,2}$ 
are gaugino masses, I find 
\begin{equation}
\delta = {\sin \theta \cos \theta \over 16 \pi^2} \left[ (3g_2^2-g_1^2) \ln 
{\tilde m_1^2 \over \mu^2} - {1 \over 4} (3g_2^2+g_1^2) \left( \ln 
{\tilde m_1^2 \over \tilde m_2^2} - {1 \over 2} \right) \right].
\end{equation}
Using $\Delta m_{32}^2 = 2.5 \times 10^{-3}$ eV$^2$ from the atmospheric 
neutrino data, this implies that
\begin{equation}
\left[ \ln {\tilde m_1^2 \over \mu^2} - 0.3 \left( \ln {\tilde m_1^2 \over 
\tilde m_2^2} - {1 \over 2} \right) \right] \sin \theta \cos \theta 
\simeq 0.535 \left( {0.4 ~{\rm eV} \over m_0} \right)^2.
\end{equation}
To the extent that the factor on the left cannot be much greater than unity, 
this means that $m_0$ cannot be much smaller than about 0.4 eV \cite{klapdor}.

In the presence of $Im \delta''$, as shown by Eq.~(30), 
\begin{equation}
U_{e3} \simeq {i Im \delta'' \over \sqrt 2 \delta},
\end{equation}
and the previous expressions for the neutrino mass eigenvalues are still 
approximately valid with the replacement of $\delta'$ by $\delta' + 
(Im \delta'')^2 /2 \delta$ and of $\delta''$ by $Re \delta''$.  There 
is also the relationship
\begin{equation}
\left[ {\Delta m_{12}^2 \over \Delta m_{32}^2} \right]^2 \simeq \left[ 
{\delta' \over \delta} + |U_{e3}|^2 \right]^2 + \left[ {Re \delta'' \over 
\delta} \right]^2.
\end{equation}
Using $\Delta m_{12}^2 \simeq 5 \times 10^{-5}$ eV$^2$ from solar neutrino 
data and $|U_{e3}| < 0.16$ from reactor neutrino data \cite{react}, I find
\begin{eqnarray}
&& Im \delta'' < 8.8 \times 10^{-4} ~(0.4~{\rm eV}/m_0)^2, \\ 
&& Re \delta'' < 7.8 \times 10^{-5} ~(0.4~{\rm eV}/m_0)^2.
\end{eqnarray}

In conclusion, recent experimental progress on neutrino oscillations points 
to a neutrino mixing matrix which can be understood in a systematic way 
\cite{ma02} in terms of an all-purpose neutrino mass matrix, i.e. Eq.~(5), 
and its simple extension, i.e. Eq.~(22), to allow for a nonzero and 
$imaginary$ $U_{e3}$, i.e. Eq.~(28).  Seven possible cases have been 
identified, each with a different prediction for $\beta \beta_{0\nu}$ decay, 
i.e. Eqs.~(15) to (21).  A specific example is that of an underlying $A_4$ 
symmetry at some high energy scale, which allows the observed Majorana 
neutrino mass matrix to be derived from radiative corrections.  It has been 
shown \cite{bmv} that this automatically leads to $\sin^2 2 \theta_{atm} = 1$ 
and a large (but not maximal) solar mixing angle.  Using neutrino oscillation 
data, and assuming radiative corrections from soft supersymmetry breaking, the 
effective mass measured in neutrinoless double beta decay is predicted to 
be not much less than 0.4 eV.

This work was supported in part by the U.~S.~Department of Energy
under Grant No.~DE-FG03-94ER40837.

\bibliographystyle{unsrt}

\end{document}